\documentclass[preprint,onecolumn,nofootinbib]{revtex4}
\usepackage[colorlinks=true,linkcolor=blue,urlcolor=blue,filecolor=black,citecolor=red,pdfstartview=FitV,pdfpagemode=None,bookmarksopen=true]{hyperref}

\usepackage{graphicx}   
\usepackage{amsmath}    
\usepackage{amsfonts}   
\usepackage{amssymb}    
\usepackage{tikz}       
\usepackage{dcolumn}    
\usepackage{xcolor}     
\usepackage{float}      
\usepackage{subfig}     
\usepackage{overpic}    
\usepackage{natbib}     

\begin{document}
\title{
  Dynamics of spontaneous scalarization of black holes with nonlinear electromagnetic fields in anti-de Sitter spacetime
}
\author{Ke-Tai Wu $^{1}$}
\email{tankwu888@gmail.com}
\author{Zi-Jun Zhong $^{1}$}
\email{zjzhong@stu2020.jnu.edu.cn}
\author{Yi Li $^{1}$}
\email{yilyu@stu2022.jnu.edu.cn}
\author{Chong-Ye Chen $^{1}$}
\email{cychen@stu2022.jnu.edu.cn}
\author{Cheng-Yong Zhang $^{1}$}
\email{zhangcy@email.jnu.edu.cn}
\author{Chao Niu $^{1}$}
\email{niuchaophy@gmail.com}
\author{Peng Liu $^{1}$}
\email{phylp@email.jnu.edu.cn}
\thanks{corresponding author}
\affiliation{
  $^1$ Department of Physics and Siyuan Laboratory, Jinan University, Guangzhou 510632, China
}

\begin{abstract}

  We investigate spontaneous scalarization in the Einstein-Born-Infeld-Scalar (EBIS) model with asymptotically AdS boundary conditions, revealing novel dynamical critical phenomena in black hole evolution. Through numerical analysis, we discover a distinctive ``flip" phenomenon where the scalar field exhibits critical transitions between different stable configurations. These transitions manifest in two forms: a single flip under variations in initial perturbation amplitude or scalar-electromagnetic coupling, and a double flip when varying black hole charge. Near critical points, the system displays universal relaxation behavior characterized by logarithmic scaling of relaxation time, $\tau \propto \ln |p - p_s|$, where $p_s$ denotes the critical initial amplitude. We demonstrate that these transitions arise from the system's approach to unstable AdS-Born-Infeld black hole configurations, which serve as separatrices between distinct stable phases. The Born-Infeld parameter plays a crucial role in this dynamics, with scalar hair vanishing in the strong nonlinearity limit. These results reveal fundamental aspects of black hole phase transitions in theories with nonlinear electromagnetic couplings and provide new insights into critical phenomena in gravitational systems.

\end{abstract}

\maketitle
\tableofcontents

\section{Introduction}
\vspace{10pt}
\label{sec:introduction}

The no-hair theorem stands as a cornerstone of general relativity, asserting that charged spinning black holes in Einstein-Maxwell theory are completely characterized by only three parameters: mass, angular momentum, and charge \cite{ruffini1971introducing,carter1971axisymmetric,israel1967event}. However, recent studies have revealed compelling challenges to this theorem through the phenomenon of spontaneous scalarization, particularly in extended scalar-tensor-Gauss-Bonnet (eSTGB) gravity \cite{doneva2018new,Fernandes:2019rez,Herdeiro:2018wub} , Einstein-Maxwell-Scalar (EMS) theories \cite{herdeiro2018spontaneous,zhang2022dynamicalEMS,luo2022dynamical,jiang2024type,niu2022dynamical,astefanesei2019einstein} and Einstein-Maxwell-Vector (EMV) theories \cite{oliveira2021spontaneous,ye2024spontaneous}. This mechanism demonstrates how initially ``bald'' black holes can develop scalar hair due to tachyonic instabilities, not only challenging the no-hair theorem but also unveiling rich dynamical behaviors. 

Furthermore, recent studies have extended the scope of spontaneous scalarization to include couplings between scalar fields and nonlinear electrodynamics, leading to new insights into the dynamics of scalarization in the presence of nonlinear electromagnetic fields. In particular, the Einstein-Born-Infeld-Scalar (EBIS) model represents a natural extension of these studies, incorporating the Born-Infeld electrodynamics as a well-motivated nonlinear theory of electromagnetism \cite{east2017superradiant,herdeiro2017dynamical,brito2020superradiance,wang2021scalarized,zhang2022spherical}. On the other hand, asymptotically AdS black hole spacetimes have attracted a large amount attention due to the value in non-trivial phenomenon and AdS/CFT correspondance. The influence of the AdS background on the black holes dynamics in the presence of nonlinear electrodynamics is a topic of great interest, as it may reveal new insights into the dynamics of scalarization and the formation of hairy black holes \cite{ashtekar1984asymptotically,stuchlik1999some,ashtekar2000asymptotically,cai2004born,fernando2006thermodynamics,myung2008thermodynamics}.

In our investigation of the EBIS model, we explore a fundamental question regarding the dynamical nature of spontaneous scalarization: How do black holes transition between different scalarized states, and what governs these transitions? This question is particularly compelling because it probes the intersection of nonlinear electrodynamics, scalar field dynamics, and black hole physics. While previous studies in EMS theories have revealed interesting dynamical behaviors including scalar field transitions \cite{chen2023time,chen2307nonlinear}, the role of nonlinear electromagnetic effects in such processes remains largely unexplored. The EBIS model, with its Born-Infeld nonlinearity, provides an ideal framework for investigating these dynamics, as it naturally interpolates between linear Maxwell theory and the strong-field regime where nonlinear effects become dominant. Understanding these transitions is crucial not only for completing our picture of spontaneous scalarization but also for revealing potential universal features in the dynamics of modified theories of gravity. Moreover, the AdS boundary conditions introduce additional richness to the system, potentially unveiling new connections between bulk dynamics and boundary physics through the AdS/CFT correspondence.

This paper is organized as follows: In Section \ref{sec:EBISmodel}, we present the theoretical framework of the EBIS model and derive the relevant field equations. Section \ref{sec:nonlinear_evolution} details our numerical methods, initial conditions, and results, including the dynamics of scalarization for various coupling functions, the dependence on Born-Infeld and cosmological parameters, and a detailed analysis of the flip phenomenon. Section \ref{flip_mechanism} delves into the analysis behind the occurrence of the flip phenomenon and explains the underlying dynamic mechanisms. Finally, Section \ref{sec:summary} discusses the implications of our findings and outlines directions for future research.

\section{Einstein-Born-Infeld-Scalar model}
\label{sec:EBISmodel}

We consider a scalar field $\phi$ non-minimally coupled to the Born-Infeld (BI) electromagnetic field $A_{\mu}$ into an asymptotically AdS spherical black hole system \cite{fernando2003charged,zh2008phases,doneva2010quasinormal,zou2014critical,li2016dyonic,tao2017testing,liang2020phase,gan2019strong} . The action for the four-dimensional Einstein-Born-Infeld-Scalar (EBIS) model in AdS spacetime is given by:
\begin{equation}\label{eq:eqact}
  S=\int d^{4}x\sqrt{-g}\left[R-2\Lambda - 2\partial_{\mu}\phi\partial^{\mu}\phi+\frac{4f(\phi)}{a}\left(1-\sqrt{1+\frac{a}{2}F^2}\right)\right],
\end{equation}
where $R$ is the Ricci scalar, $\Lambda$ is the cosmological constant, and $a$ is the Born-Infeld parameter characterizing the strength of nonlinear effects. The electromagnetic field strength tensor $F_{\mu\nu} = \partial_\mu A_\nu - \partial_\nu A_\mu$ appears in the Lorentz invariant kinetic term $F^2 = F_{\mu\nu}F^{\mu\nu}$. The function $f(\phi)$ describes the non-minimal coupling between the scalar field $\phi$ and the electromagnetic field $A_\mu$. In this paper, we employ geometric units where $16\pi G = 1$.

The Born-Infeld term in the action, $\frac{4f(\phi)}{a}(1 - \sqrt{1 + \frac{a}{2}F^2})$, exhibits distinct behaviors at different limits. As $a \to 0$, this term reduces to $-f(\phi)F^2$, corresponding to the EMS theory. In contrast, as $a \to \infty$, the BI term effectively vanishes, reducing the action to the Hilbert term along with the kinetic term of a massless scalar field. For finite $a$, the term introduces significant nonlinear corrections, especially as the field strength approaches the critical value $\sqrt{2/a}$. This nonlinear structure is particularly notable because it allows for the regularization of the electromagnetic self-energy of a point charge, thereby addressing a fundamental limitation in classical electrodynamics, as originally highlighted by Born and Infeld in their foundational work \cite{born1934foundations}.

The equations of motion for system \eqref{eq:eqact} are:
\begin{align}
  R_{\mu\nu} - \frac{1}{2}Rg_{\mu\nu} + \Lambda g_{\mu\nu}                              & = \frac{1}{2}T_{\mu\nu},       \label{eq:eqmot}               \\ 
  \partial_{\mu}\left[\frac{\sqrt{-g}f(\phi)F^{\mu\nu}}{\sqrt{1+\frac{a}{2}F^2}}\right] & = 0,                \label{eq:maxwell}                        \\
  \frac{1}{\sqrt{-g}}\partial_{\mu}(\sqrt{-g}\partial^{\mu}\phi)                        & = -\frac{f'(\phi)}{a}\left(1-\sqrt{1+\frac{a}{2}F^2}\right),
\end{align}
where the $T_{\mu\nu}$ is energy-momentum tensor:
\begin{equation}\label{eq:eqEMT}
  T_{\mu\nu} = 4\left(\partial_{\mu}\phi\partial_{\nu}\phi - \frac{1}{2}g_{\mu\nu}\partial_{\rho}\phi\partial^{\rho}\phi\right) + 4f(\phi)\left[\frac{1}{a}\left(1-\sqrt{1+\frac{a}{2}F^2}\right)g_{\mu\nu} + \frac{F_{\mu\rho}F_{\nu}{}^{\rho}}{\sqrt{1+\frac{a}{2}F^2}}\right].
\end{equation}

Following \cite{martel2001regular,chesler2014numerical}, we adopt the ingoing Eddington-Finkelstein coordinates to investigate the fully nonlinear dynamics of a spherical black hole in an asymptotically AdS spacetime. The metric ansatz reads
\begin{equation}
  \label{eq:eqmetric}
  ds^2 = -\alpha(t,r)dt^2 + 2dtdr + \zeta(t,r)^2(d\theta^2 + \sin^2\theta d\varphi^2),
\end{equation}
where $\alpha$ and $\zeta$ are metric functions dependent on $(t,r)$. The apparent horizon radius $r_h$ is defined by the condition $g^{\mu\nu}\partial_{\mu}\zeta\partial_{\nu}\zeta = 0$. Accordingly, the thermodynamic entropy of the black hole is proportional to the area of this apparent horizon, $V_h = 4\pi\zeta(r_h,t)^2$, and the irreducible mass of the black hole is defined as $M_h \equiv \sqrt{V_h/(4\pi)} = \zeta(r_h,t)$. On the other hand, the ansatz of gauge field is $A_{\mu}dx^{\mu} = A(t,r)dt$. The Born-Infeld equations (as derived from Eq. \eqref{eq:maxwell}) yield a conserved quantity along the $r$-direction, which is associated with the black hole charge $Q$. By further deducing from this conserved quantity, one arrives at the equation
\[
  \partial_r A = \frac{Q}{\zeta^2 f}.
\]

To simplify the equations of motion, we introduce two auxiliary variables:
\begin{align}
  S & \equiv \partial_t\zeta + \frac{1}{2}\alpha\partial_r\zeta, \label{eq:eqauxvarS} \\
  N & \equiv \partial_t\phi + \frac{1}{2}\alpha\partial_r\phi. \label{eq:eqauxvarN}
\end{align}
Note that at the apparent horizon, the condition $g^{\mu\nu}\partial_{\mu}\zeta\partial_{\nu}\zeta = 0$ implies $S = 0$. substituting these auxiliary variables, the Einstein equations can be written as:
\begin{align}
   & \partial_t S = \frac{S \partial_r {\alpha} - \alpha \partial_r S}{2} - \zeta N^2, \label{eq:partial_t_S}                                                                                                                                                                                                \\
   & \partial_r ^2 \zeta = -\zeta (\partial_r \phi)^2, \label{eq:partial_r2_zeta}                                                                                                                                                                                                                            \\
   & \partial_r S = \frac{1 - 2 S  \partial_r \zeta }{2 \zeta} - \frac{\Lambda \zeta}{2} + \frac{f(\phi) \zeta \left(1 - \sqrt{\frac{f(\phi)^2\zeta^4}{a Q^2 + f(\phi)^2\zeta^4}} \right)}{a} -\frac{Q^2 \sqrt{\frac{f(\phi^2)\zeta^4}{a Q^2 + f(\phi)^2\zeta^4}}}{f(\phi)\zeta ^ 3}, \label{eq:partial_r_S} \\
   & \partial_r ^2 \alpha = - 4 N \partial_r \phi  + \frac{4 S \partial_r \zeta  - 2}{\zeta^2}+ \frac{4 Q^2 \sqrt{\frac{f(\phi)^2\zeta^4}{a Q^2 + f(\phi)^2\zeta^4}}}{f(\phi)\zeta ^ 4}. \label{eq:partial_r2_alpha}
\end{align}
The scalar field equation becomes:
\begin{equation}\label{eq:partial_r_N}
  \begin{aligned}
    \partial_r N = - \frac{N \partial_r \zeta +S \partial_r \phi}{\zeta} - \frac{1 - \sqrt{\frac{f(\phi)^2\zeta^4}{a Q^2 + f(\phi)^2\zeta^4}} }{2a} \frac{df}{d\phi}.
  \end{aligned}
\end{equation}
These equations consistently incorporate the Born-Infeld electrodynamic term, reducing to the case in EMS model as $a \to 0$.

By solving the aforementioned equations, we can numerically simulate the evolution of the black hole system. First, Eq.~\eqref{eq:partial_r2_zeta} is employed to determine the initial profile of $\zeta(r, t=0)$ by specifying an initial scalar field $\phi(r, 0)$. With $\zeta(r, 0)$ and $\phi(r, 0)$ known, Eq.~\eqref{eq:partial_r_S} is solved to obtain $S(r, 0)$. Subsequently, $N(r, 0)$ is computed from Eq.~\eqref{eq:partial_r_N}. Once $S$, $\zeta$, and $N$ are established, Eq.~\eqref{eq:partial_r2_alpha} is solved to determine $\alpha(r, 0)$. This procedure exclusively involves linear solvers, ensuring high precision for the spatial solutions on each individual spacetime foliation. For the time evolution of $\phi$, we employ the Runge-Kutta method, updating $\phi$ via Eq.~\eqref{eq:eqauxvarN}.

\subsection{Perturbation Analysis and Tachyonic Instability}

In the probe limit, the scalar field can be treated as the perturbation $\delta \phi$ within the background of a BI black hole, and the perturbative equation of motion can be derived as
\begin{align}
   & \frac{\partial_{\mu}(\sqrt{- g}\partial^{\mu} \delta \phi)}{\sqrt{- g}} = \mu_{\text{eff}}^{2} \delta \phi, \\
   & \mu_{\text{eff}}^{2} = -\ddot f(0)\frac{1- r^{2} \sqrt{r^{4} +a Q^{2}}}{a Q^{2}} ,\label{eq:eqmiueff},
\end{align}
where $\mu_{\text{eff}}^{2}$ represents the effective square mass of the perturbative scalar field and $\ddot{f}(0)$ denotes the $\frac{d^2 f(\phi)}{d \phi^2}\Big{|}_{\phi=0}$. If $\mu_{\text{eff}}^{2}<0$, the perturbation $\delta \phi$ will undergo an exponential growth due to the tachyonic instability, leading to the spontaneous scalarization of the black hole. This mechanism typically leads to the formation of scalar hair around the bald black hole. Moreover, based on \eqref{eq:eqmiueff}, this condition further requires $\ddot f(0)>0$.

The coupling function $ f(\phi) = e^{\beta \phi^2}$ satisfies $\dot f(0) = 0$ while maintain $\ddot f(0)>0$ as long as $\beta>0$. Therefore, in this work, we focus on this exponential coupling function to investigate the spontaneous scalarization in the EBIS model.

\subsection{Boundary Conditions and Initial Conditions of AdS Spacetime}

The appropriate initial and boundary conditions are crucial for a stable and accurate numerical simulation of the EBIS model. We begin by considering the asymptotic behavior of the variables $\phi$, $\alpha$, $\zeta$, $S$, and $N$ at spatial infinity. By substituting asymptotic expansions of these variables into the Einstein field equations, we obtain the following expressions:
\begin{align}
   & \phi(t,r)    = \frac{\phi_3 (t)}{r^3} + \frac{3}{8 \Lambda r^4} \left(\frac{Q^2 f'(0)}{f(0)^2} - 8 \phi_3 ' (t)\right) + O\left(r^{-5}\right),                                         \\
   & \zeta(t,r)  = r - \frac{3 \phi_3^2 (t)}{10 r^5} + \frac{3 \phi_3 (t)}{14 \Lambda r^6} \left(\frac{Q^2 f'(0)}{f(0)^2} - 8 \phi_3 ' (t)\right) + O\left(r^{-7}\right),                   \\
   & S(t,r)       = - \frac{\Lambda}{6} r^2 + \frac{1}{2} - \frac{M}{r} + \frac{Q^2}{2 f(0) r^2} - \frac{3 \Lambda}{20 r^4} \phi_3^2 (t) + O\left(r^{-5}\right),                            \\
   & N(t,r)       = \frac{\Lambda \phi_3 (t)}{2 r^2} + \frac{1}{r^3} \left(\frac{Q^2 f'(0)}{4 f(0)^2} -  \phi_3 ' (t)\right) + \frac{3}{2\Lambda r^4} \phi_3 '' (t) + O\left(r^{-5}\right), \\
   & \alpha(t,r)  = - \frac{\Lambda}{3} r^2 + 1 - \frac{2M}{r} + \frac{Q^2}{f(0) r^2} + \frac{\Lambda}{5 r^4} \phi_3^2(t) + O\left(r^{-5}\right).
\end{align}
Here, $f'(\phi) = \frac{df(\phi)}{d\phi}\Big|_{r\to\infty}$, $\phi_3'(t) = \frac{d\phi_3(t)}{dt}$, $M$ represents the ADM mass  \cite{abbott1982stability} , $Q$ the black hole charge, and $\Lambda$ the cosmological constant. For simplicity, we set $M=1$ and $\Lambda=-3$, and maintain the geometric units throughout. The function $\lambda(t)$ represents a gauge freedom in the radial coordinate, which allows us to fix the apparent horizon radius during the evolution.

These asymptotic behaviors will lead to divergence in numerical simulation as $r\to\infty$. To fix this, we redefine the variables $(\phi, \zeta, S, N, \alpha)$ into new variables $(\varphi, \gamma, s, n, \alpha_{1})$ as follows:
\begin{align}
   & \phi(t,r)=\frac{\varphi(t,r)}{r^{3}},                              \\
   & \zeta(t,r)=r+ \lambda(t) +\frac{\gamma(t,r)}{r^{3}},               \\
   & S(t,r) =\frac{1}{2}+\frac{(r+\lambda(t))^{2}}{2}-\frac{s(t,r)}{r}, \\
   & N(t,r)=-\frac{3\varphi(t,r)}{2L^{2}r^{2}}+\frac{n(t,r)}{r^{3}},    \\
   & \alpha (t,r)=\frac{(r+\lambda(t))^{2}}{L^{2}}+1+\alpha_{1}(t,r),
\end{align}
which ensures all new variables are finite at the AdS boundary, while satisfying the boundary conditions.

In our numerical framework, the gauge degree of freedom, denoted by $\lambda(t)$, is crucial for controlling the radial position of the apparent horizon during the evolution process, as highlighted in prior studies \cite{chesler2014numerical,zhang2022dynamicalEMS}. Specifically, we define $r_{\text{hi}} = r_h + \lambda(t)$, where $r_h$ denotes the apparent horizon's radius in a fixed radial coordinate. By evolving $\lambda(t)$ dynamically, we fix the apparent horizon at $r_h = 1$ while allowing the physical radius to evolve via the gauge freedom inherent in $\lambda(t)$. The dynamical evolution of $\lambda(t)$ can be determined by the boundary condition at the AdS boundary,
\begin{equation}
  \partial_{t}\lambda = -\frac{1}{2}\lim_{r\to\infty}\alpha_{1}(t,r).
\end{equation}
Thanks to the nature of the equations, and given that $S = 0$ is itself the condition for a Killing horizon, along with the order of derivatives involved in the variables, we can extract further boundary conditions from the time derivative of $ S $. This, in turn, provides the boundary condition for $ \alpha $, expressed as:
\begin{align}
   & S(r_{h},\lambda)=0,                                                                                                \\
   & \partial_{t}S\Big|_{r_{h}}=0\Leftrightarrow\alpha(r_{h},\lambda)=-\frac{2\zeta N^{2}}{\partial{r}S}\Big|_{r_{h}}.
\end{align}
For the remaining variables $\zeta$ and $N$, we employ natural boundary conditions.

To fully characterize the system, we must specify both the boundary and initial conditions. For the latter, we initialize the scalar field with a Gaussian wave packet, given by
\begin{equation}
  \begin{aligned}
    \phi(t=0,r)=p e^{-(\frac{r-r_{0}}{w})^{2}}.
  \end{aligned}
\end{equation}
where $p$, $r_{0}$, and $w$ represent the initial amplitude, center, and width of the Gaussian wave packet, respectively. We set $w=r_h$ and $r_0=4r_h$, with $r_h=1$ for convenience. As a result, the initial amplitude $p$ becomes the key parameter in our study of the time evolution.

Moreover, thanks to the simplicity of fixing horizon radius, we can employ a compactification for radial coordinate, mapping $r$ to $z = \frac{1}{r}$. This transformation changes the range of radial coordinate from $r \in [r_{h}, \infty)$ to $z \in [0, z_h]$, where $z_h = \frac{1}{r_h}=1$.

In this work, we employ Chebyshev-Lobatto collocation to discretize the radial coordinate $z$ and the fourth-order Runge-Kutta (RK4) method for time evolution, balancing accuracy and computational efficiency. The complete numerical procedure can be summarized as follows:
\begin{enumerate}
  \item Solve the spatial equations to obtain $(\gamma, s, n, \alpha_{1})$ based on the current scalar field configuration $\varphi$ and gauge $\lambda$ for one time slice.
  \item Use the RK4 method to compute the updated values of the dynamical variables $ (\varphi, \lambda) $.
  \item Sum the update increments to obtain the new values of $ \varphi $ and $ \lambda $, which then serve as the initial conditions for the next time step.
  \item Repeat steps 1-3 for each subsequent time step.
\end{enumerate}
This approach allows us to effectively simulate the dynamics of the asymptotically AdS black hole while maintaining numerical stability and accuracy.

\section{Fully Nonlinear Dynamical Evolution}
\label{sec:nonlinear_evolution}

In this section, we present a detailed investigation of spontaneous scalarization in charged black holes within the Einstein Born-Infeld Scalarization model. Through numerical evolution of the fully nonlinear field equations, we uncover rich dynamical behaviors including critical phenomena and phase transitions. Our analysis reveals how the scalar field dynamics depends on key parameters: the initial perturbation amplitude ($p$), black hole charge ($Q$), scalar-electromagnetic coupling ($\beta$), and Born-Infeld parameter ($a$). We demonstrate the existence of distinct evolution patterns, critical points exhibiting both single and double flip transitions, and the crucial role of nonlinear electrodynamics in maintaining scalar hair.

\subsection{Types of Evolution Process}\label{subseq:Typical_Time_Evolution}

\begin{figure}[H]
  \centering
  \includegraphics[width=0.60\textwidth]{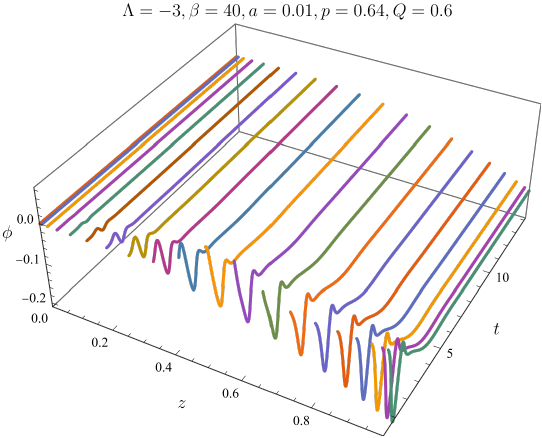}
  \captionsetup{justification=raggedright}
  \caption{Time evolution behaviors of the scalar field. }
  \label{fig:exam}
\end{figure}

In Fig.~\ref{fig:exam}, we illustrate the evolution of the scalar field at various positions $z$. Our analysis reveals that, although the amplitudes of the scalar field vary at different positions $z$, they exhibit a qualitatively similar evolutionary pattern. Therefore, in our subsequent investigations, we will focus on the scalar field values at horizon ($z=1$), as these are deemed representative results for presentation.

\begin{figure}[H]
  \centering
  \includegraphics[width=0.60\textwidth]{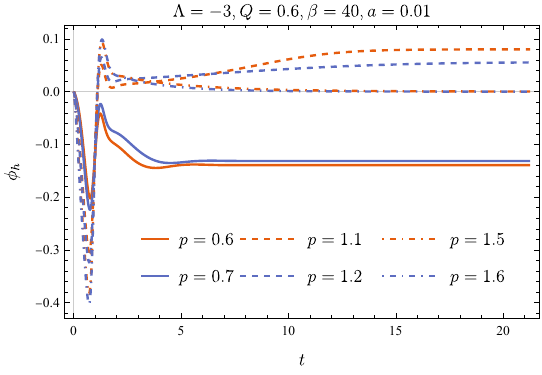}
  \captionsetup{justification=raggedright}
  \caption{Typical time evolution behaviors of the scalar field with different $p$.}
  \label{fig:normal}
\end{figure}

In the EBIS model, we identify three characteristic patterns of time evolution for the scalar field.
\begin{itemize}
  \item \textbf{Type I:} Rapid oscillations converging to a negative stable value.
  \item \textbf{Type II:} Rapid oscillations converging to a positive stable value.
  \item \textbf{Type III:} Rapid oscillations followed by complete dissipation.
\end{itemize}
Fig.~\ref{fig:normal} illustrates these three behaviors, with the solid line representing Type I, the dashed line corresponding to Type II, and the dotted line depicting Type III. With other parameters fixed, we observe Type I behavior for $p = 0.6, 0.7$, Type II for $p = 1.1, 1.2$, and Type III for $p = 1.5, 1.6$. These values of the Gaussian wave packet parameter $p$ produce qualitatively different scalar field dynamics, including both scalarization and descalarization phenomena. While we demonstrate these patterns by varying $p$, similar behavioral transitions emerge when varying other parameters such as $Q$ and $\beta$, indicating the generality of these three fundamental patterns in the system.

\label{Typical_Time_Evolution}

The systematic variation of $p$ reveals a clear transition pattern in the final state of the scalar field $\phi$. As $p$ increases from lower values, we observe that the stabilized scalar field transitions from negative values (Type I) through positive values (Type II), and ultimately approaches zero (Type III), corresponding to a complete descalarization or ``bald" state. This behavior suggests the existence of a critical value $p_s$ where the scalar field changes sign during stabilization. This transition point may serve as a bifurcation parameter, marking the boundary between distinct phases of the system's long-term behavior.

\begin{figure}[H]
  \centering
  \subfloat[]{\includegraphics[width=0.45\textwidth]{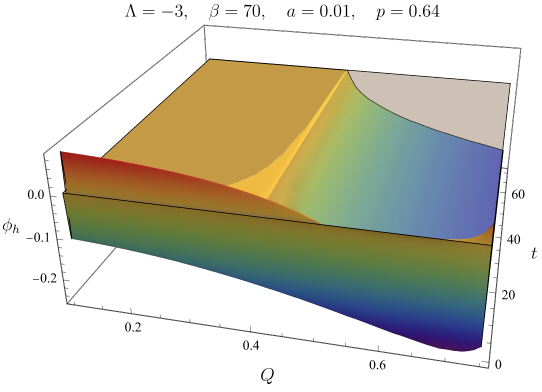}}\hspace{1pt}
  \subfloat[]{\includegraphics[width=0.45\textwidth]{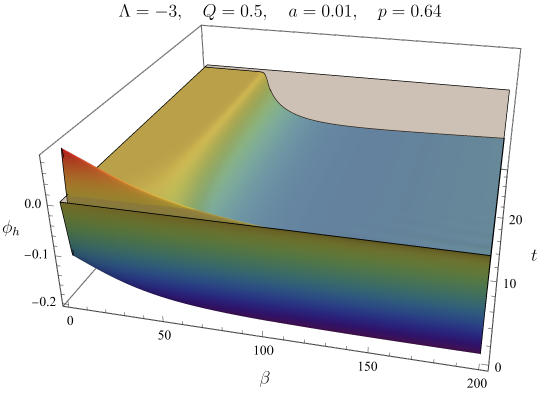}}\hspace{1pt}
  \caption{Typical time evolution behavior of the scalar field, as a function of $Q$ and $\beta$. }
  \label{fig:evooQ1}
\end{figure}

To explore how black hole evolution depends on charge $ Q $ and coupling constant $ \beta $, we conducted full time evolution simulations across different values of $ Q $ or $ \beta $, while keeping $ p $ and $ a $ fixed (see Fig. \ref{fig:evooQ1}).

Fig.~\ref{fig:evooQ1} presents two 3D plots showing the scalar field amplitude at the horizon $ \phi_h $ as a function of evolution time and the parameters $(Q, \beta)$. A semi-transparent gray surface at $ \phi_h = 0 $ is included in both plots to clearly delineate regions where the scalar field vanishes, revealing the descalarization dynamics.

Our numerical analysis reveals that for small values of either the coupling constant $ \beta $ or the charge $ Q $, the scalar field rapidly dissipates during the evolution. This behavior suggests that the stable formation of hairy black holes requires both $ Q $ and $ \beta $ to exceed critical threshold values. These findings highlight the crucial role of the electromagnetic field in facilitating the process of spontaneous scalarization within this system. Moreover, the transitions among the three evolutionary patterns illustrated in Fig.\ref{fig:normal} can also be achieved by varying parameters such as $Q$ and $\beta$, as shown in Fig.\ref{fig:evooQ}.

\subsection{Dynamical Critical Phenomena}\label{Special_Flip_Evolution}

In this EBIS model, the evolution of the scalar field follows a distinct two-stage process: an initial period characterized by rapid oscillations and growth, followed by stabilization at a finite stable value. A particularly intriguing feature emerges when varying the wave packet amplitude $p$: the system displays a distinctive bifurcation phenomenon where the stable value of the scalar field undergoes sign reversal with only small variations in $p$. 

\begin{figure}[H]
  \centering
  \includegraphics[width=0.70\textwidth]{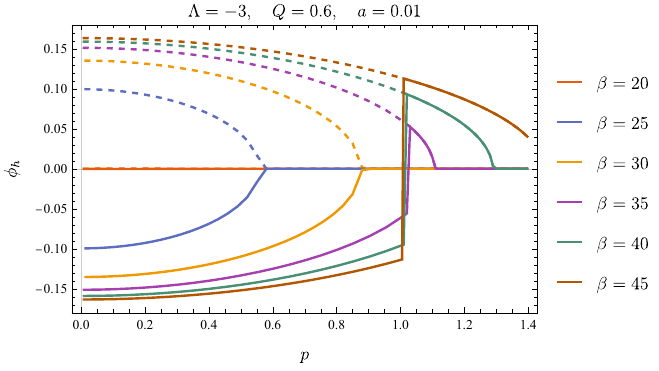}
  \captionsetup{justification=raggedright}
  \caption{
    Under different initial conditions, the value of the scalar field $\phi_{h}$ at the black hole horizon changes with the wave packet initial amplitude $p$. The dotted line represents the absolute value of the solid line.
  }
  \label{fig:evoabsp}
\end{figure}

Fig.~\ref{fig:evoabsp} illustrates the dependence of the final scalar field value at the horizon, $\phi_{h}$, on $p$ for different choices of the coupling constant $\beta$. For $\beta = 25$ and $\beta = 30$, $\phi_{h}$ remains negative and approaches zero as $p$ increases, leading to a bald black hole in the final state. However, for larger coupling constants, e.g., $\beta = 35$, $\beta = 40$, and $\beta = 45$, the behavior changes significantly. As $p$ increases, $\phi_{h}$ initially rises slowly from a negative value, then abruptly transitions to a positive value when crossing a critical point $p_s$, before decreasing towards zero. This sharp transition indicates the presence of a bifurcation. More rigorously, the sign-flipping behavior near the critical point $p_s$ is described by the limit
\begin{equation}
  \lim_{p \to p_s^-} \phi(p) = -\lim_{p \to p_s^+} \phi(p),
\end{equation}
which reflects that, when crossing the critical amplitude $p_s$, the scalar field at the horizon flips its sign while maintaining its magnitude.

This sign-flipping phenomenon is further validated through analysis of $|\phi_h|$ in the region where $\phi_h < 0$ (dotted line, Fig.~\ref{fig:evoabsp}). The smooth continuity of $|\phi_h|$ across the transition point $p_s$ demonstrates that the bifurcation induces only a sign change in $\phi_h$ while preserving its magnitude. This behavior indicates that the sign flip does not introduce a discontinuity in the spacetime geometry.

\begin{figure}[H]
  \centering
  \includegraphics[width=0.50\textwidth]{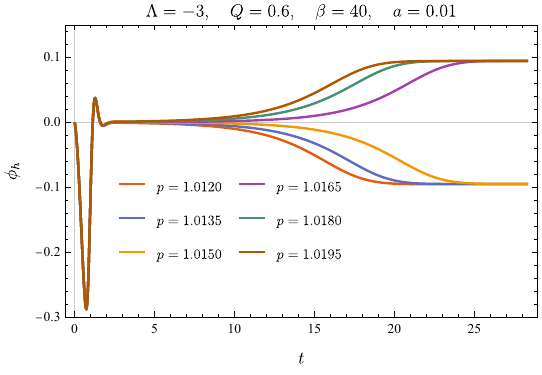}
  \captionsetup{justification=raggedright}
  \caption{
    The scalar field exhibits critical relaxation behavior near the flip critical point. Where $\phi\to 0$ is the critical point 
  }
  \label{fig:evonearptwo}
\end{figure}

Our analysis demonstrates that the flip phenomenon, while leaving the final stabilized geometry unaffected, has a significant impact on the dynamical behavior of the system. As shown in Fig.~\ref{fig:evonearptwo}, we examine the time evolution of the scalar field for $p$ values on either side of the flip critical point $p_s$. The scalar field will experience a phase of relaxation, characterized by exponential growth from values approaching zero. Notably, the relaxation time increases as $p$ approaches $p_s$, indicating critical slowing down. At precisely $p = p_s$, the system dissipates the initially finite scalar field and evolves into a unstable  AdS-BI black hole configuration. It is important to highlight that $p_s$ acts as an unstable fixed point in the context of renormalization group (RG) flow, where any deviation from $p_s$ drives the system away from this critical point.

\begin{figure}[H]
  \centering
  \subfloat[]{\includegraphics[width=0.42\textwidth]{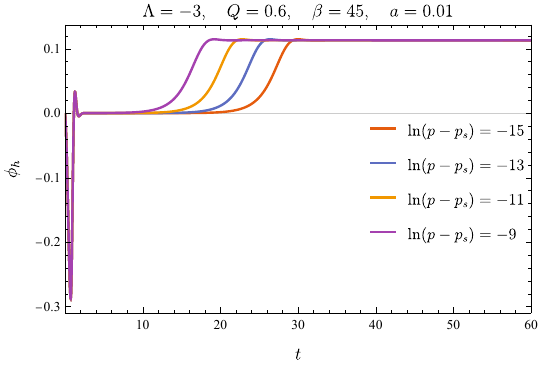}\label{fig:evoln_a}}\hspace{1pt}
  \subfloat[]{\includegraphics[width=0.42\textwidth]{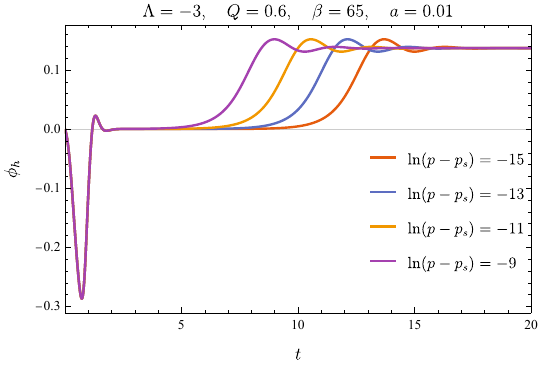}\label{fig:evoln_b}}\hspace{1pt}
  \subfloat[]{\includegraphics[width=0.42\textwidth]{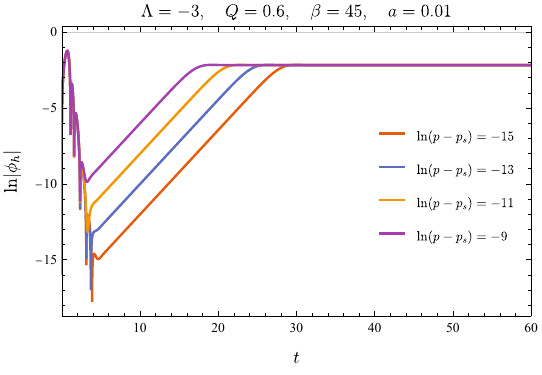}\label{fig:evoln_e}}\hspace{1pt}
  \subfloat[]{\includegraphics[width=0.42\textwidth]{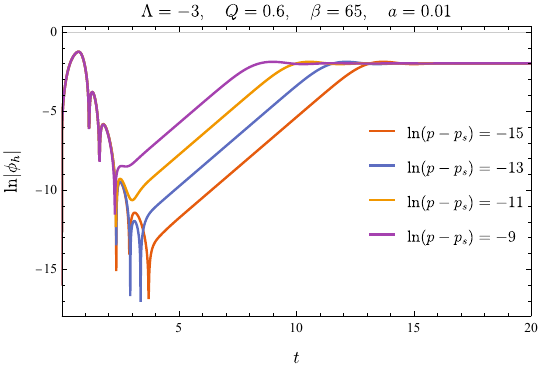}\label{fig:evoln_f}}\vspace{1pt}
  \subfloat[]{\includegraphics[width=0.42\textwidth]{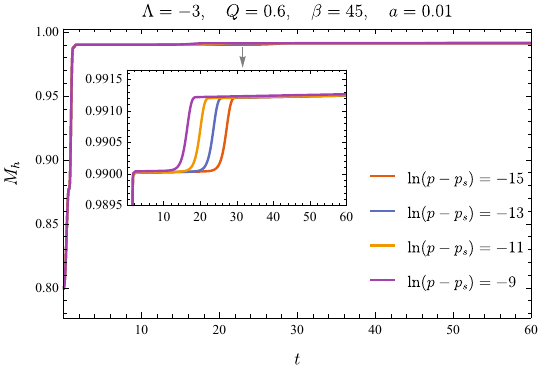}\label{fig:evoln_c}}\hspace{1pt}
  \subfloat[]{\includegraphics[width=0.42\textwidth]{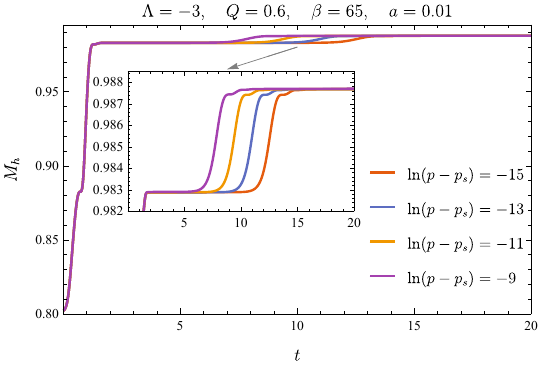}\label{fig:evoln_d}}\hspace{1pt}
  \captionsetup{justification=raggedright}
  \caption{Evolution of key parameters near the singular point $p_s$ for two values of $\beta$. Left column: $\beta=45$ ($p_s=1.008205$); right column: $\beta=65$ ($p_s=0.987547$). From top to bottom: scalar field $\phi_h$, logarithm $\ln|\phi_h|$, and irreducible mass $M_h$.}
  \label{fig:evoln}
\end{figure}

To rigorously establish the critical nature of the flip phenomenon, we focus on the behavior of three key physical quantities near the transition point: the scalar field value at the horizon $\phi_h$, the logarithmic scaling of the scalar field $\ln|\phi_h|$, and the irreducible mass, $M_h$. Fig.~\ref{fig:evoln} illustrates the time evolution of these quantities for two representative coupling functions, $f(\phi) = pe^{45\phi^2}$ and $f(\phi) = pe^{65\phi^2}$. Specifically, $\phi_h(t)$ captures the time-dependent dynamics of the scalar field, $\ln|\phi_h(t)|$ reveals the presence of critical scaling behavior, and, $M_h(t)$ tracks the evolution of the black hole's mass. To parameterize the system's proximity to the critical point, we define the control parameter $\Delta p \equiv p - p_s$, where $p_s$ is the critical coupling determined through the bisection method, corresponding to the threshold at which the scalar field undergoes a sign change. The time evolution of $\phi_h$ for various values of $\Delta p$ (depicted in different colors in Fig.~\ref{fig:evoln}) systematically reveal how the system responds as it approaches the critical point.

As $p$ approaches $p_s$, the scalar field $\phi_h$ asymptotically approaches zero, accompanied by a critical slowing down in the system's relaxation dynamics, as evidenced by the behavior of $M_h$. Precisely at $ p = p_s $, the scalar field vanishes, and the black hole configuration corresponds to an AdS-BI black hole. Fig.~\ref{fig:evoln_a} and \ref{fig:evoln_b} demonstrate that $\phi_h$ increases from zero in an exponential manner, a behavior further corroborated by the linear growth of $\ln|\phi_h|$ over time shown in Fig.~\ref{fig:evoln_e} and \ref{fig:evoln_f}. This exponential growth is indicative of a critical point where $\phi$ approaches zero, causing the background geometry to converge to that of the original Born-Infeld black hole. The relaxation time increases as $p$ nears $p_s$, reflecting the critical slowing down phenomenon. Additionally, the relaxation behavior observed in the black hole mass $M_h$ (see Fig.~\ref{fig:evoln_c} and \ref{fig:evoln_d}) further substantiates the criticality of the system. Collectively, these observations affirm that the flip point $p_s$ embodies a critical state characterized by distinctive relaxation dynamics of both the scalar field and the black hole.

Although the above figures clearly demonstrate critical relaxation behavior near the flip point, concluding that this represents the dynamical critical phenomenon requires more careful examination. In particular, a distinctive feature of dynamical critical phenomanon is the scaling of the relaxation time $\tau$ with the perturbation parameter $\Delta p = |p - p_s|$, typically following a power-law relationship $\tau \sim \Delta p^\alpha$, where $\alpha$ is the scaling exponent. By investigating this scaling behavior, we aim to determine whether the observed critical relaxation indeed corresponds to a dynamical critical phenomenon.

To rigorously represent the relaxation process, we find that it can be well fit with the exponential decay model given in Eq. \eqref{filp_phi_b}. Specifically, we model the dynamical behavior of $\phi_h$ near the flip point as:  
\begin{equation}  
  \label{filp_phi_b}
  \phi_h(t) = e^{-kt + a},  
\end{equation}
where $k$ and $a$ are constants, determined by fitting the data. The relaxation time $\tau$ is then defined as:  
\begin{equation}  
  \tau = \frac{a}{k}.  
  \label{critical}
\end{equation}
This relaxation time $\tau$ characterizes the timescale for the scalar field to evolve to its stable value, and its consistency with the observed behavior validates the robustness of the model. The timescale $\tau$ thus serves as a critical metric for understanding the stabilization dynamics of $\phi_h$.

\begin{figure}[H]
  \centering
  \includegraphics[width=0.64\textwidth]{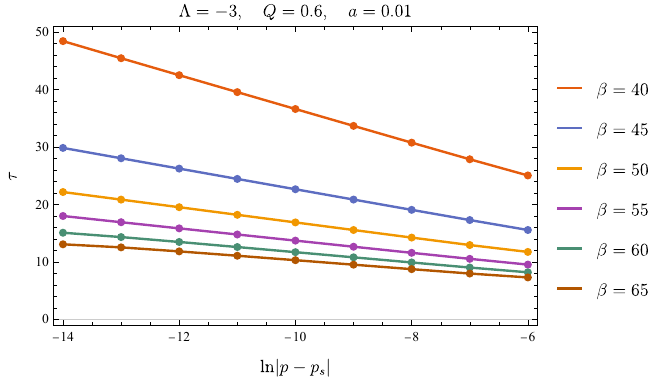}
  \captionsetup{justification=raggedright}
  \caption{Under different $\beta$, linear behavior of relaxation time $\tau$ near the flip point, lines of different colors represent various values of $\beta$, while the horizontal axis denotes $\ln(p - p_s)$. Since changes in $\beta$ affect the value of $p_s$, each line corresponds to a distinct value of $p_s$.}
  \label{fig:evolineart}
\end{figure}

Next, we analyze the scaling behavior of $\tau$ as the system approaches the critical point. Fig.~ \ref{fig:evolineart} demonstrates a linear relationship between $\ln |p - p_s|$ and $\tau$,
\begin{equation}
  \tau \propto \ln |p - p_s|.
\end{equation}
This logarithmic relationship suggests that the relaxation time grows exponentially as the system nears the flip point. Such exponential growth is a characteristic feature of systems near criticality and provides a strong indication of the presence of a critical point. This scaling behavior is commonly observed in dynamical critical phenomenon, providing compelling evidence that the flip phenomenon indeed corresponds to a dynamical critical phenomenon. The scaling exponent $\alpha$ may depend on system-specific parameters, such as $\beta$ and $a$.

\subsection{Single Filp vs Double Flip Dynamics}\label{subseq:Single_Double_Flip}

While our previous analysis focused on the flip phenomenon driven by parameter $p$, the dynamical critical phenomenon exhibits distinctly different behaviors when driven by other parameters. A systematic investigation of the parameter space reveals an intriguing phenomenon: the existence of both single and double flips in the scalar field evolution.

\begin{figure}[H]
  \centering
  \subfloat[]{\includegraphics[width=0.42\textwidth]{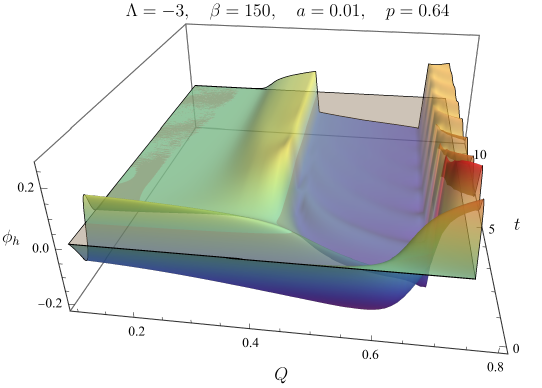}}\hspace{45pt}
  \subfloat[]{\includegraphics[width=0.42\textwidth]{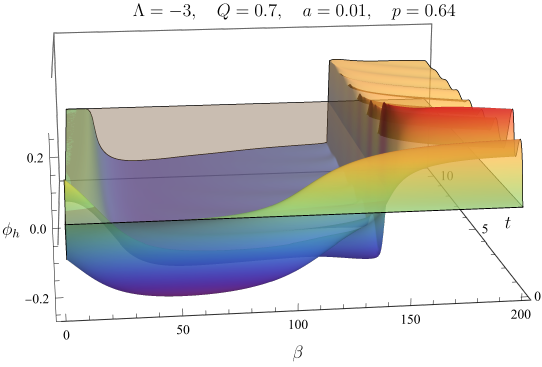}}\hspace{15pt}
  \caption{Three-dimensional visualization of the scalar field evolution. (a) Evolution of $\phi_h$ with charge $Q$, exhibiting double flip behavior. (b) Evolution with coupling constant $\beta$, showing single flip behavior. A reference plane at $\phi_h = 0$ is included to highlight the transitions.}
  \label{fig:evooQ}
\end{figure}

Fig.~\ref{fig:evooQ} illustrates this fundamental difference through the evolution of the horizon scalar value $\phi_h$ under variations of charge $Q$ and coupling constant $\beta$. The most striking feature appears in the charge-driven evolution (Fig.~\ref{fig:evooQ}(a)), where $\phi_h$ undergoes two successive transitions: first from positive to negative, and then back to positive as $Q$ increases. This double flip behavior stands in sharp contrast to the single flip we observed in the $p$-driven evolution (Fig.~\ref{fig:evoabsp}).

The occurrence of a double flip indicates that this dynamical system exhibits a rich criticality, characterized by complex coupling relationships with the system parameters, and revealing a non-monotonic phase geometric structure.

This distinction between single and double flips suggests that the dynamical phase transitions in this system can exhibit qualitatively different behaviors depending on which parameter drives the evolution. Understanding the conditions that determine the number of flips and their physical origins remains an important question for future investigation.

\subsection{Role of Born-Infeld Parameter in Black Hole Dynamics}

Having established the distinction between single and double flips in the parameter space, we now turn our attention to another crucial parameter: the Born-Infeld (BI) factor $a$. This parameter is particularly interesting as it connects different theoretical limits: as $a \to 0$, the system reduces to Einstein-Maxwell-Scalar (EMS) theory, while as $a \to \infty$, it approaches Einstein-Scalar (ES) theory without electromagnetic fields. This feature allows us to systematically study how nonlinear electrodynamics affects the spontaneous scalarization.

\begin{figure}[H]
  \centering
  \includegraphics[width=0.64\textwidth]{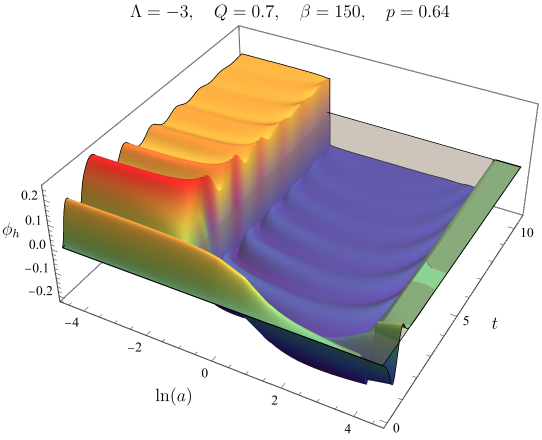}
  \captionsetup{justification=raggedright}
  \caption{Time evolution of the horizon scalar field $\phi_h$ as a function of the Born-Infeld parameter $a$. The horizontal axis uses a logarithmic scale to capture both EMS ($a \to 0$) and ES ($a \to \infty$) limits.}
  \label{fig:evoa}
\end{figure}

Fig.~\ref{fig:evoa} presents the evolution of the horizon scalar value $\phi_h$ as a function of $\ln(a)$. The logarithmic scale enables us to explore both the EMS limit (small $a$) and the ES limit (large $a$) within a single plot.

Our analysis reveals two significant features in the $a$-dependent evolution. First, increasing $a$ triggers a dynamical critical phenomenon, similar to those observed with other parameters. Second, and more remarkably, further increase in $a$ leads to descalarization, where the scalar hair completely vanishes in the final state. 

These observations highlight the unique role of nonlinear electrodynamics in spontaneous scalarization. The BI parameter $a$ emerges as a control parameter comparable to the coupling constant $\beta$ in its ability to induce phase transitions. Moreover, the absence of scalar hair in the ES limit (large $a$) demonstrates that electromagnetic fields are essential for maintaining the scalarized state, acting both through their non-minimal coupling to the scalar field and through nonlinear corrections. This dual influence provides deeper insight into the mechanisms underlying spontaneous scalarization in this system.

The flip phenomenon in the EBIS model exhibits a rich dynamical structure characterized by critical dependence on the control parameters $p$, $Q$, $\beta$, and $a$. The universality of this behavior across parameter space suggests a fundamental mechanism governing the transitions between distinct black hole configurations. We proceed to examine the underlying dynamics of this critical phenomenon.

\section{Analysis of Flip Mechanisms}
\label{flip_mechanism}

Having observed both the flip phenomenon and various relaxation behaviors in previous sections, we now focus on understanding the underlying mechanism of the flip itself. The key insight is that the flip occurs when the system approaches an unstable fixed point corresponding to a bald AdS-BI black hole. This critical configuration plays a central role in determining the system's evolution and the associated relaxation dynamics.

\subsection{Critical Solutions and Scalar Evolution}

\begin{figure}
  \centering
  \subfloat[]{\includegraphics[height=0.32\textwidth]{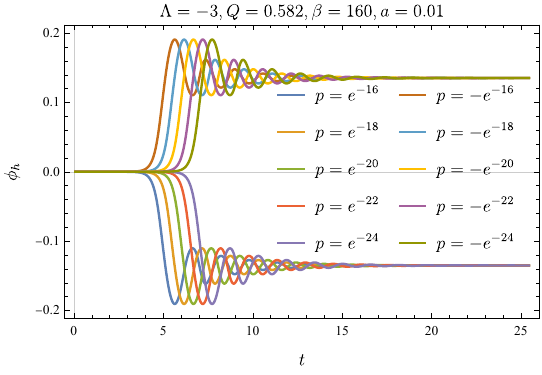}\label{fig:pos_flip}}
  \subfloat[]{\includegraphics[height=0.32\textwidth]{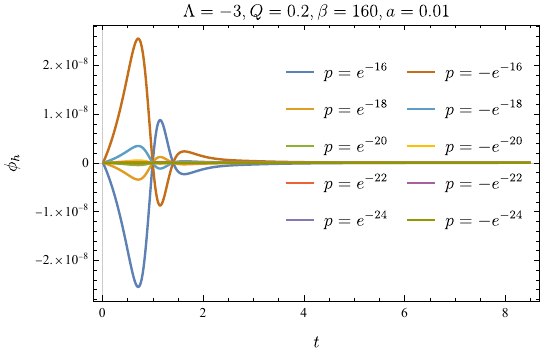}\label{fig:neg_flip}}
  \captionsetup{justification=raggedright}
  \caption{
    Evolution of scalar field under different parameters. Panel (a) shows the evolution with parameters $Q = 0.5828$, $\beta = 160$, and $a = 0.01$, demonstrating both positive and negative departures leading to scalarized solutions. Panel (b) depicts the evolution with parameters $Q = 0.2$, $\beta = 160$, and $a = 0.01$, showing descalarization where the scalar field $\phi$ eventually vanishes.
  }
  \label{fig:critical_flip}
\end{figure}

Consider a carefully chosen set of parameters where the system represents a bald AdS-BI black hole at its critical point. When we introduce a small perturbation $\delta\phi$ to this critical solution, the system's evolution exhibits a remarkable property: depending on the sign of the perturbation, the scalar field will evolve toward either positive or negative stable values. This behavior is illustrated in Fig.~\ref{fig:critical_flip}.

This bifurcation behavior is a direct consequence of the linearity of the perturbation equations near the critical point. When $p > 0$, the system evolves toward a positive scalar hair configuration; when $p < 0$, it evolves toward a negative one. The magnitude of the final scalar field value remains the same in both cases, reflecting the symmetry of the underlying equations.

\subsection{Relaxation Dynamics and QNM Analysis}

The relaxation time $\tau$ observed during the flip exhibits a characteristic logarithmic dependence on the initial perturbation amplitude:
\begin{equation}\label{eq:relaxation_time}
  \tau \propto \ln |p|
\end{equation}
This scaling can be understood through quasi-normal mode (QNM) analysis. For small perturbations, the system's evolution is dominated by the leading QNM. Near the critical point, the perturbative scalar field follows:
\begin{equation}
  \delta\phi(t) \sim p e^{\omega_I t} = e^{\ln |p| + \omega_I t}
\end{equation}
where $\omega_I$ is the imaginary part of the leading QNM frequency. The time required for the perturbation to grow to a significant amplitude naturally scales with $\ln |p|$, explaining the observed relaxation behavior \eqref{eq:relaxation_time}.

\begin{figure}
  \centering
  \includegraphics[width=0.5\textwidth]{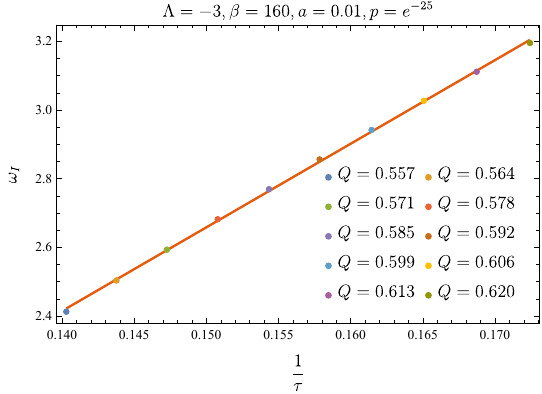}
  \captionsetup{justification=raggedright}
  \caption{The relationship between inverse relaxation time ($1/\tau$) and imaginary frequency ($\omega_I$) for different charge ($Q$) at a fixed perturbation magnitude of $p=e^{-25}$.}
  \label{fig:qnmvstime}
\end{figure}

We performed numerical calculations of the quasi-normal modes to verify this relationship. Fig. \ref{fig:qnmvstime} shows the correlation between the imaginary part of the QNM frequency ($\omega_I$) and the relaxation time ($\tau$), demonstrating a clear linear dependence. This direct proportionality between $\omega_I$ and $\tau$ provides strong quantitative support for our analytical understanding of the relaxation dynamics.

\subsection{Phase Structure and Stability}

\begin{figure}[H]
  \centering
  \includegraphics[width=0.6\textwidth]{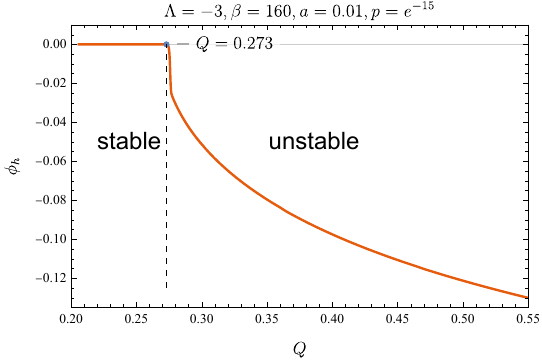}
  \caption{Phase diagram showing $\phi_h$ vs $Q$ for small perturbations. The left side (zero $\phi_h$) represents the stable branch of bald AdS-BI black holes, while the right side (nonzero $\phi_h$) indicates the unstable bald black holes where scalar hair can grow. This structure remains largely unchanged for different small values of $p$.}
  \label{fig:phase_diagram}
\end{figure}

The relationship between critical solutions and stable configurations can be visualized through a phase diagram. Fig.~\ref{fig:phase_diagram} shows the value of the scalar field at the horizon ($\phi_h$) as a function of charge $Q$.

This phase diagram reveals several important features:
\begin{itemize}
  \item The unstable branch (dashed line) represents critical solutions where the flip can occur
  \item Any solution on this branch, under small perturbations, evolves toward one of the stable branches
  \item The structure remains robust for different small values of $p$, though the detailed evolution may vary
\end{itemize}

It is important to note that while we focus on the simple picture presented above, the actual dynamics can be more complex. The system's evolution depends on multiple parameters ($Q$, $\beta$, $p$) and even the initial perturbation configuration. This creates a rich structure of intersecting hypersurfaces in the full parameter space.

The flip mechanism can be understood as a consequence of the system approaching unstable critical solutions corresponding to bald AdS-BI black holes. These critical points serve as separatrices between positive and negative scalar hair configurations. The relaxation dynamics near these points is governed by quasi-normal modes, leading to the characteristic logarithmic scaling of relaxation time with perturbation amplitude. This framework provides a unified understanding of both the flip phenomenon and its associated relaxation behavior.

While previous investigations identified the flip phenomenon within the EMS model primarily in the context of external quenching \cite{chen2023time}, our findings demonstrate that this phenomenon also emerges through initial bulk perturbations. Specifically, the flip phenomenon manifests when the initial black hole evolves into an unstable AdS-BI black hole following perturbations, with its occurrence governed by the system's fundamental parameters ($p$, $Q$, $\beta$, and $a$). This behavior is systematically mapped across various parameter spaces, as evidenced in Figs.~\ref{fig:evoabsp}, \ref{fig:evooQ}, and \ref{fig:evoa}. Furthermore, our analysis reveals that the EBIS system exhibits similar flip dynamics, indicating that this phenomenon, along with the associated scalarization, persists in the nonlinear electrodynamics. These results collectively suggest that the flip phenomenon represents a universal dynamical feature in black hole systems.

\section{Summary And Discussion}
\label{sec:summary}

In this work, we have conducted a comprehensive investigation of spontaneous scalarization in the Einstein-Born-Infeld-Scalar model, revealing rich dynamical behaviors and critical phenomena that emerge from the interplay between nonlinear electrodynamics and scalar fields in asymptotically AdS spacetimes. Our findings not only extend our understanding of black hole phase transitions but also illuminate fundamental aspects of critical phenomena in gravitational systems.

The most significant discovery is the identification of a robust dynamical critical phenomenon characterized by distinct flip transitions in the scalar field evolution. Through detailed numerical analysis, we demonstrated that these transitions manifest in multiple ways across the parameter space, exhibiting both single and double flip behaviors depending on the controlling parameters. The single flip, observed when varying the initial perturbation amplitude $p$ or the coupling constant $\beta$, represents a direct transition between opposite-signed scalar hair configurations. In contrast, the double flip, emerging under variations in the black hole charge $Q$, reveals a more intricate phase structure where the system undergoes successive transitions through intermediate states. This distinction in flip behaviors suggests that the underlying phase space possesses a rich geometric structure, with different parameters accessing distinct paths through this space.

The critical nature of these transitions is firmly established through our analysis of the relaxation dynamics. The observed logarithmic scaling of relaxation time, $\tau \propto \ln |p - p_s|$, provides strong evidence for critical behavior analogous to classical phase transitions. This scaling relationship, combined with our quasi-normal mode analysis, demonstrates that the flip phenomenon represents a genuine dynamical critical point rather than a mere instability. The universality of this scaling across different parameter regimes suggests that it reflects a fundamental feature of spontaneous scalarization in nonlinear electromagnetic environments.

A particularly illuminating aspect of our study is the role of the Born-Infeld parameter $a$ in modulating the scalarization process. By connecting the limits of standard Einstein-Maxwell-Scalar theory ($a \to 0$) and pure Einstein-Scalar theory ($a \to \infty$), we have shown how nonlinear electromagnetic effects fundamentally influence the stability and dynamics of scalar hair. The observation that scalar hair vanishes in the large-$a$ limit demonstrates that electromagnetic nonlinearity plays a crucial role in maintaining scalarized states, suggesting a deep connection between field nonlinearity and the stability of hairy black hole configurations.

Our results also shed new light on the relationship between spontaneous scalarization and black hole stability. The identification of unstable AdS-Born-Infeld black holes as critical points separating different stable phases provides a unified framework for understanding both the flip phenomenon and the general process of spontaneous scalarization. This framework suggests that the flip transitions we observe are manifestations of a more general principle governing phase transitions in modified theories of gravity.

Moreover, several promising directions emerge from this work. First, the existence of double flip transitions raises questions about the possibility of even more complex transition patterns in systems with additional degrees of freedom or different types of nonlinear couplings. Second, the role of AdS boundary conditions in shaping these transitions deserves further investigation, particularly in light of potential holographic interpretations through the AdS/CFT correspondence. Finally, the universal aspects of the critical behavior we observed suggest the possibility of a broader classification scheme for dynamical phase transitions in gravitational systems.

\section*{Acknowledgments}

We thank Qian Chen for helpful discussions. Peng Liu would like to thank Yun-Ha Zha and Yi-Er Liu for kind encouragement during this work. This work is supported by the Natural Science Foundation of China under Grant No. 12475054 and No.12375048.

 \bibliographystyle{unsrtnat}
\bibliography{misc/references.bib}

\end{document}